\documentclass[review, sort&compress]{elsarticle}
\usepackage{epsfig,amsmath,graphicx,amssymb,overpic}
\usepackage{bm}
\usepackage{mathrsfs}
\usepackage{graphicx}
\usepackage{epsfig}
\usepackage{dcolumn}
\usepackage{bm}
\usepackage{amsmath,amssymb,amsthm}
\usepackage[colorlinks=true,linkcolor=blue]{hyperref}
\usepackage{subfigure}
\usepackage{booktabs}
\usepackage[mathscr]{euscript}
\usepackage{ulem}
\usepackage{color}
\usepackage{geometry}  

\usepackage{CJK}
\begin{document}

\title{Multi-component decompositions, linear superpositions and new nonlinear integrable coupled KdV-type systems}
\author[rvt1]{Xiazhi Hao\corref{cor1}}%
\author[rvt2]{S. Y. Lou}
\cortext[cor1]{Corresponding author. Email: haoxiazhi2008@163.com (Xiazhi Hao), lousenyue@nbu.edu.cn(S. Y. Lou) }

\address[rvt1]{College of Science, Zhejiang University of Technology, Hangzhou, 310014, China}
\address[rvt2]{School of Physical Science and Technology, Ningbo University, Ningbo, 315211, China}

\begin{abstract}	
	The existence of decompositions of the nonlinear integrable systems not only permits us to establish so-called linear superposition solutions, but also to derive new nonlinear integrable coupled systems. 
	Restricting our attention to the single component decompositions of the potential BKP hierarchy, we obtain that suitable linear superpositions of some decomposition solutions still satisfy the same equations.
In parallel, successful attempts are made  by multi-component decompositions of the potential BKP hierarchy to construct linear superposition solutions and new nonlinear integrable coupled KdV-type systems that cannot be decoupled by a change of dependent variables.
\end{abstract}

\maketitle

\section{Introduction}

A common way of dealing with high-dimensional nonlinear systems is to split these systems into several low-dimensional ones, which are easier to treat with the available tools \cite{sbib22,ebib13}. We handled some types of nonlinear partial differential equations, the potential B-type Kadomtsev-Petviashvili (pBKP) equations and heavenly equations, for instance, by making single component decompositions which convert them into equations we already knew how to solve. Some decompositions reveal a rare property that their suitable linear superpositions are still in fact new solutions to the same equations. The linear superposition principle implies, in particular, that in the process of nonlinear evolution every single solution propagates almost independently of the other solutions even though they may collide in physical space for a certain period and the exact solution equals the sum of these particular single solutions \cite{sbib35,sbib41,wbib22}. 

This paper is a direct continuation of our previous work about the decompositions and special linear superpositions \cite{xbib24} which leads us to seek multi-component decompositions of the pBKP hierarchy, correspondingly, a feasible way to construct integrable couplings is furnished.
	The BKP equation known as (2+1)-dimensional Sawada-Kotera (SK) equation \cite{ybib17,rbib20} 
	\begin{equation}
	u_{xt}+(u_{x4}+15uu_{x2}+15u^3-15uv-5u_{xy})_{xx}-5u_{yy}=0,\ v_x=u_y, \label{BKP}
	\end{equation}
	where $u_x=\partial_x u, u_{x2}=\partial_x^2u, u_{x3}=\partial_x^3u,\dots,$
	was discovered by Konopelchenko and Dubrovsky \cite{bbib2}.
	We convert BKP equation \eqref{BKP} to a potential form
	\begin{equation}
	w_{xt}=5w_{yy}-(w_{x5}+15w_xw_{x3}+15w_x^3-15w_xw_y-5w_{xxy})_{x},\label{PBKP}
	\end{equation}
	 which can be represented as compatibility condition in the Lax form 
	\begin{eqnarray}
	&&\psi_y-\psi_{x3}-3w_x\psi_x=0,\label{Lx}\\
	&&\psi_t-9\psi_{x5}-45 w_x \psi_{x3}-45 w_{xx}\psi_{xx} -15(2 w_{x3}+3w_x^2+w_y)\psi_x=0,\label{Lt}
	\end{eqnarray}
	by setting $u=w_x$ that is simple enough to be significantly more convenient to use.

Back in paper \cite{xbib24}, the pBKP hierarchy constructed in terms of the mastersymmetry approach is the following class of commuting flows \cite{bbib7,ybib16}
	\begin{equation}\label{PBKPH}
	w_t=K_{2n-1}=\frac{1}{3\cdot 5^n n!}K_{[,]}^{n}y^{n},\quad n=1,\ 2,\ \ldots,\ \infty,
	\end{equation}
	where the commutate operator $K_{[,]}$ is defined as
	\begin{equation}\label{Kf'}
	K_{[,]}f\equiv K'f-f'K\equiv \lim_{\epsilon\rightarrow0}\frac{\rm d}{\rm d\epsilon} \left[K(w+\epsilon f)-f(w+\epsilon K)\right]
	\end{equation}
	for arbitrary $f$.
	The first five equations of this hierarchy are listed
\begin{equation}\label{K1}
w_{t_1}=K_{1}=\frac{1}{15}K_{[,]}y=w_x,
\end{equation}
\begin{equation}\label{K3}
w_{t_3}=K_{3}=\frac{1}{150}K_{[,]}^2y^2=3w_y,
\end{equation}
\begin{equation}\label{K5}
w_{t_5}=K_{5}=\frac{1}{2250}K_{[,]}^3y^3=5\partial_x^{-1}w_{y2}-w_{x5}-15w_xw_{x3}-15w_x^3 +5w_{yx2}+15w_x w_y=K,
\end{equation}
\begin{eqnarray}\label{K7}
w_{t_7}&=&K_{7}=\frac{1}{45000}K_{[,]}^4y^4\\
&=&-w_{x7} -21 w_{x3}^2-21 w_{x2}w_{x4} -21 w_x w_{x5} +21\partial_x^{-1}(w_xw_{yy})+7\partial_x^{-2}w_{y3}+42 w_{x2}w_{xy}\nonumber\\
&&-42w_{x3}w_y +42\partial_x^{-1}(w_{x4}w_y)-63\partial_x^{-1}(w_x^2w_{xy}) -21w_xw_{yx2}-63w_xw_{xx}^2 -126w_x^2w_{x3} +21w_{xy2}\nonumber\\
&&+21 w_x\partial_x^{-1}w_{y2}-63w_x^4+63/2w_y^2,\nonumber
\end{eqnarray}
\begin{eqnarray}\label{K9}
w_{t_9}&=&K_{9}=\frac{1}{1125000}K_{[,]}^5y^5\\
&=&9\partial_x^{-2}\{\partial_x^{-1}w_{y4}-w_{yx8}
+6 w_{x2y3}-270 w_{x} w_{x2}^2 w_{y}+3 w_{x5y2}+36 w_{yx2}^2+9 w_{x}w_{x3y2} +9 w_{x}w_{y3}\nonumber\\
&&+3(10 w_{xy2}-105 w_{x2}w_{xy}-111 w_{x}w_{yx2}-f) w_{x3} -9(39 w_{x}w_{x2}+5 w_{xy}) w_{yx3}-18 w_xw_{yx6}\nonumber\\
&& +27 w_xw_{xy}^2+15 w_{xy2}w_y-3 w_{x7} w_{y}
-9(21 w_xw_{xy}+10 w_{x2} w_y-w_{y2}+7 w_{yx3}) w_{x4}  +36 w_{x2}w_{x2y2}
\nonumber\\
&&-45 w_{x2}w_{yx5}+9 w_{x2}\partial_x^{-1}w_{y3}-18 w_{x6}w_{xy} +27 w_{xy}w_{y2} +9[2(5 w_y-27 w_x^2) w_{xy} +3 w_xw_{y2}] w_{x2}\nonumber\\
&&+3(5 w_y-27 w_x^2-20 w_{x3}) w_{yx4}-45(w_xw_y+w_{yx2}) w_{x5}
-63 w_{x3}^2 w_y\nonumber\\
&&+9 w_{yx2}(\partial_x^{-1}w_{y2}-12 w{x}^3+5 w_xw_y-27 w_{x2}^2)\},\ \nonumber\\
f_x&=&6(15w_xw_{x2} -2w_{xy}-w_{x4})w_y+54w_x^2w_{xy}-\partial_x^{-1}w_{y3}-3(w_xw_{y})_y.\nonumber
\end{eqnarray}

For $n=3$, Equation \eqref{K5} is exactly the pBKP equation \eqref{PBKP} with $t_5=t$. All the equations in \eqref{PBKPH} are the compatibility conditions of the following linear systems
\begin{eqnarray}
&&L\psi=0,\quad L=\partial_y-\partial_{3x}-3w_x\partial_x,\\
&&\psi_{t_{2n-1}}=A_n\psi,
\end{eqnarray}
where the Lax operators are the polynomials in $\partial_x$ such that \begin{eqnarray}
3K_{2n-1,x}\partial_x=[L,A_n]
\end{eqnarray}
identically. 
All $A_n$ are generated in the same way as $K_{2n-1}$.
The first several Lax operators are listed as follows
\begin{eqnarray*}
&&A_1=\partial_x,\\
&&A_2=3\partial_y,\\
&&A_3=9\partial_{x}^5+45 w_x \partial_{x}^3+45 w_{xx}\partial_{x}^2 +15(2 w_{x3}+3w_x^2+w_y)\partial_x.
\end{eqnarray*}

In the next part of the paper, we review briefly the single component decompositions and linear superposition solutions of the pBKP equations \eqref{K5}-\eqref{K9}.
Then, we generalize in Section 3 the decomposition to the multi-component cases and introduce nonlinear integrable coupled  korteweg-de vries (KdV)-type systems that possess higher order symmetries. Finally, Section 4 is a conclusion.

\section{Single component decompositions and linear superposition solutions of the pBKP hierarchy}

Before going into the contents and details of our paper, below we remind the reader the single component decompositions of the pBKP hierarchy. More details can be found in \cite{xbib24}.

\subsection{Decompositions and linear superposition solutions of the fifth-order pBKP equation \eqref{K5}}
\noindent
\textbf{Proposition 1.}\label{pro1}
	If $w_1,\ w_2,\ w_3,\ w_4,\ w_5$ and $w_6$ are the solutions of the following decomposed systems
	\begin{eqnarray}
	&&\left\{\begin{array}{l}
	\displaystyle{w_{1y}=(\Phi_1+c) w_{1x}+c_1},\
	\Phi_1\equiv \partial_x^2+4w_{1x}-2\partial_x^{-1}w_{1x2}, \vspace{2mm}\\
	\displaystyle{w_{1t}=(9\Phi_1^2+15c \Phi_1+5(c^2+3c_1))w_{1x}},
	\end{array}\right. \label{DC51}\\
	&&\left\{\begin{array}{l}
	\displaystyle{w_{2y}=\Phi_2w_{2x}+c_1},\ \Phi_2\equiv \partial_x^2+2w_{2x}-\partial_x^{-1}w_{2x2},\vspace{2mm}\\
	\displaystyle{w_{2t}=(9\Phi_2^2+15c_1)w_{2x}},
	\end{array}\right. \label{DC52}\\
	&&\left\{\begin{array}{l}
	\displaystyle{w_{3y}=-\frac12( \Phi_3-2c) w_{3x}+c_1,\
		\Phi_3\equiv \partial_x^2+4w_{3x}-2\partial_x^{-1}w_{3x2},} \vspace{2mm}\\
	\displaystyle{w_{3t}=\bigg(5c^2+15c_1-\frac94 \Phi_3^2\bigg)w_{3x}},
	\end{array}\right. \label{DC53}\\
	&&\left\{\begin{array}{l}
	\displaystyle{{w_{4y}=w_{4x3}-\frac34\frac{w_{4x2}^2}{W^2}}+\frac32W^4-c_1= \Phi_4w_{4x}+c_1,\ W^2\equiv w_{4x}+c}, \vspace{2mm}\\
	\displaystyle{w_{4t}=9\Phi_4^2w_{4x}+15c_1w_{4x},\ \Phi_4\equiv \partial_x^{-1}{W}\partial_x^2W^{-1}\partial_x +W^2+\partial_x^{-1}W^2\partial_x,}
	\end{array}\right. \label{DC5JM}\\
	&&\left\{\begin{array}{l}
	\displaystyle{w_{5y}=\frac14(4\Phi_5^2+6c\Phi_5+3c^2)w_{5x},\ \Phi_5\equiv \partial_x+\frac12 w_5+\frac12 w_{5x}\partial_x^{-1}},\vspace{2mm}\\
	\displaystyle{w_{5t}=\frac9{16}(16\Phi_5^4+40c\Phi_5^3+40c^2\Phi_5^2+20c^3\Phi_5+5c^4)w_{5x}},
	\end{array}\right. \label{DCSTO}\\
	&&\left\{\begin{array}{l}
	\displaystyle{w_{6y}=cw_{6x}+c_1},\vspace{2mm}\\
	\displaystyle{w_{6t}=-w_{6x5}+5(c-3w_{6x})w_{6x3}+15cw_{6x}^2 -15w_{6x}^3+5(c^2+3c_1)w_{6x},}
	\end{array}\right. \label{DCSK}
	\end{eqnarray}
	then $w_1,\ w_2,\ w_3,\ w_4,\ w_5$ and $w_6$ are all solutions of the pBKP equation \eqref{K5}. 
	
The pBKP equation \eqref{K5} may be written as the compatibility condition between two classical KdV flows \eqref{DC51}-\eqref{DC53}, coupled Svinolupov–Sokolov equations \eqref{DC5JM}, coupled Sharma–Tasso–Olver equations \eqref{DCSTO} and coupled SK equations \eqref{DCSK}. Some solutions of the pBKP equation may consequently be derived from solutions of these systems. Decompositions \eqref{DC51}-\eqref{DC52} also allow us to establish linear superposition solutions of the pBKP equation \eqref{K5}. Thus, we have the following proposition which can be verified by a straightforward calculation.
\\
	\textbf{Proposition 2.}\label{pro2}
		Suppose that $w_1,\ w_2,\ w_3,\ w_4,\ w_5$ and $w_6$ are solutions of the pBKP equation \eqref{K5} with the conditions
		\begin{eqnarray}\begin{split}\label{w16}
		&w_{1y}=(\Phi_1+c) w_{1x},\
		w_{1t}=(9\Phi_1^2+15c \Phi_1+5c^2)w_{1x},\ \Phi_1\equiv \partial_x^2+4w_{1x}-2\partial_x^{-1}w_{1x2}, \\
		&w_{2y}=(\Phi_2-c) w_{2x},\
		w_{2t}=(9\Phi_2^2-15c \Phi_2+5c^2)w_{2x}, \ \Phi_2\equiv \partial_x^2+4w_{2x}-2\partial_x^{-1}w_{2x2},\\
		&w_{3y}=\Phi_3 w_{3x},\
		w_{3t}=9\Phi_3^2w_{3x}, \ \Phi_3\equiv \partial_x^2+4w_{3x}-2\partial_x^{-1}w_{3x2}, \\
		&w_{4y}=\Phi_4w_{4x},\ w_{4t}=9\Phi_4^2w_{4x}, \ \Phi_4\equiv \partial_x^2+2w_{4x}-\partial_x^{-1}w_{4x2},\\
		&w_{5y}=\Phi_5w_{5x}-\frac{c^2}6,\ w_{5t}=(9\Phi_5^2-\frac{5c^2}2)w_{5x}, \ \Phi_5\equiv \partial_x^2+2w_{5x}-\partial_x^{-1}w_{5x2},\\
		&w_{6y}=\Phi_6w_{6x}-\frac{c^2}6,\ w_{6t}=(9\Phi_6^2-\frac{5c^2}2)w_{6x}, \ \Phi_6\equiv \partial_x^2+2w_{6x}-\partial_x^{-1}w_{6x2},
		\end{split}\end{eqnarray}
		then, the linear superpositions 
		\begin{eqnarray}\label{w79}
		&&w_7=w_1+w_2,\ w_8=w_3+\frac12w_4,\ w_9=\frac12(w_5+w_6)
		\end{eqnarray} 
		are at the same time the solutions of the pBKP equation \eqref{K5}.

\subsection{Decompositions and linear superposition solutions of the seventh-order pBKP equation \eqref{K7}}
For the seventh-order pBKP equation \eqref{K7}, decompositions can be found in a similar way. We just list the result in the following proposition.		\\
\textbf{Proposition 3.}\label{pro3}
	The functions $w_i,\ i=1,\ 2,\ \ldots,\ 6$ satisfying the following decomposition systems
	\begin{eqnarray}
	&&\left\{\begin{array}{l}
	\displaystyle{w_{1y}=(\Phi_1+c) w_{1x}+c_1},\vspace{2mm}\\
	\displaystyle{w_{1t}=[27\Phi_1^3+63c \Phi_1^2 +(42c^2+63c_1)\Phi_1+7c(c^2+9c_1)]w_{1x}},
	\end{array}\right. \label{DC71}\\
	&&\left\{\begin{array}{l}
	\displaystyle{w_{2y}=\Phi_2w_{2x}+c_1},\vspace{2mm}\\
	\displaystyle{w_{2t}=(27\Phi_2^3+63c_1\Phi_2)w_{2x}},
	\end{array}\right. \label{DC72}\\
	&&\left\{\begin{array}{l}
	\displaystyle{w_{3y}=-\frac12( \Phi_3-2c)w_{3x}+c_1},\vspace{2mm}\\
	\displaystyle{w_{3t}=\bigg[\frac{27}8 \Phi_3^3-\frac{63}4 c \Phi_3^2+\frac72(3c^2-9c_1)\Phi_3+7c(c^2+9c_1)\bigg]w_{3x}},
	\end{array}\right. \label{DC73}\\
	&&\left\{\begin{array}{l}
	\displaystyle{w_{4y}=\Phi_4w_{4x}+c_1},\vspace{2mm}\\
	\displaystyle{{w_{4t}}=27\Phi_4^3w_{4x}+63c_1\Phi_4w_{4x}},
	\end{array}\right. \label{DC74}\\
	&&\left\{\begin{array}{l}
	\displaystyle{w_{5y}=\frac14(4\Phi_5^2+6c\Phi_5+3c^2)w_{5x}},\vspace{2mm}\\
	\displaystyle{w_{5t}=\frac{27}{64}(64\Phi_5^6+224c\Phi_5^5 +336c^2\Phi_5^4
		+280 c^3\Phi_5^3+140c^4\Phi_5^2+42c^5\Phi_5+7c^6)w_{5x}},
	\end{array}\right. \label{DC75}\\
	&&\left\{\begin{array}{l}
	\displaystyle{w_{6y}=cw_{6x}+c_1},\vspace{2mm}\\
	\displaystyle{w_{6t}=84c_1cw_{6x}+7c^3w_{6x}+21(3w_{6x}^2+w_{6x3})c^2-21(w_{6x}^3 +w_{6x}w_{6x3}-w_{6x2}^2)c }\vspace{2mm}\\
	\displaystyle{\qquad -63w_{6x}(w_{6x}^3+2w_{6x}w_{6x3}+w_{6x2}^2)-21(w_{6x}w_{6x5}+w_{6x2}w_{6x4}+w_{6x3}^2)-w_{6x7}}
	\end{array}\right. \label{DC76}
	\end{eqnarray}
	are all solutions of the seventh-order pBKP equation \eqref{K7} with $t_7=t$.\\
	\textbf{Proposition 4.}\label{pro4}
		If $w_1,\ w_2,\ w_3,\ w_4,\ w_5$ and $w_6$ are solutions of the seventh-order pBKP equation \eqref{K7} with conditions
		\begin{eqnarray}\begin{split}\label{w7pbkp}
		&w_{1y}=(\Phi_1+c) w_{1x},\ w_{1t}=(27\Phi_1^3+63c \Phi_1^2 +42c^2\Phi_1+7c^3)w_{1x}-c,\\
		&w_{2y}=(\Phi_2-c) w_{2x},\ w_{2t}=(27\Phi_2^3-63c \Phi_2^2 +42c^2\Phi_2-7c^3)w_{2x}-c_1,\\
		&w_{3y}=\Phi_3 w_{3x},\ w_{3t}=27\Phi_3^3w_{3x}-c,\\
		&w_{4y}=\Phi_4w_{4x},\ w_{4t}=27\Phi_4^3w_{4x}-c_1,\\
		&w_{5y}=\Phi_5w_{5x}+c_1,\ w_{5t}=9(3\Phi_5^3+7c_1\Phi_5)w_{5x}-c,\\
		&w_{6y}=\Phi_6w_{6x}+c_1,\ w_{6t}=9(3\Phi_6^3+7c_1\Phi_6)w_{6x}-c_1,\\
		\end{split}\end{eqnarray}
		then the special combinations corresponding to 
		\begin{equation}
		w_7=w_1+w_2,\ w_8=w_3+\frac12w_4,\ w_9=\frac12(w_5+w_6)\label{SK7w789}
		\end{equation}
		are also solutions of the seventh-order pBKP equation \eqref{K7}.

\subsection{Decompositions and linear superposition solutions of the ninth-order pBKP equation \eqref{K9}}

In the same way, we directly write down the decomposition proposition for the ninth-order pBKP equation \eqref{K9} without detailed verifications. \\
\textbf{Proposition 5.} \label{pro5}
	The functions $w_i,\ i=1,\ \ldots,\ 6$ provide solutions of the ninth-order pBKP equation \eqref{K9} with $t_9=t$ if they satisfy
	\begin{eqnarray}
	&&\left\{\begin{array}{l}
	\displaystyle{w_{1y}=(\Phi_1+c) w_{1x}+c_1},\quad \mu\equiv \frac92(2c^4+54c_1c^2+45c_1^2),\vspace{2mm}\\
	\displaystyle{w_{1t}=[81\Phi_1^4+243c\Phi_1^3+243(c^2+c_1)\Phi_1^2
		+18c(5c^2+27c_1) \Phi_1 +\mu] w_{1x}},
	\end{array}\right. \label{DC91}\\
	&&\left\{\begin{array}{l}
	\displaystyle{w_{2y}=\Phi_2w_{2x}+c_1},\vspace{2mm}\\
	\displaystyle{w_{2t}=\bigg(81\Phi_2^4+243c_1\Phi_2^2+\frac{405}2c_1^2\bigg)w_{2x}-c},
	\end{array}\right. \label{DC92}\\
	&&\left\{\begin{array}{l}
	\displaystyle{w_{3y}=-\frac12(\Phi_3-2c) w_{3x}+c_1},\vspace{2mm}\\
	\displaystyle{w_{3t}=\bigg[\frac{81}{16}\Phi_3^4-\frac{81(2c^2+3c_1)}{4} \Phi_3^2+\frac{9c(8c^2-9c_1)}{2} \Phi_3+\mu\bigg]w_{3x}},
	\end{array}\right. \label{DC93}\\
	&&\left\{\begin{array}{l}
	\displaystyle{w_{4y}=\Phi_4w_{4x}+c_1,\ }\vspace{2mm}\\
	\displaystyle{{w_{4t}}=81\Phi_4^4w_{4x}+243c_1\Phi_4^2w_{4x}+\frac{405c_1^2}{2}w_{4x}},
	\end{array}\right. \label{DC94}\\
	&&\left\{
	\begin{array}{l}
	\displaystyle{w_{5y}=\frac14(4\Phi_5^2+6c\Phi_5+3c^2)w_{5x},}\vspace{2mm}\\
	\displaystyle{w_{5t}=\frac{81}{256}(256\Phi_6^8+1152c\Phi_5^7+2304c^2\Phi_5^6+2688c^3\Phi_5^5
		+2016c^4\Phi_5^4 +1008c^5\Phi_5^3}\vspace{2mm}\\ 
	\displaystyle{\qquad +336c^6\Phi_5^2+72c^7\Phi_5+9c^8)w_{5x}},
	\end{array}\right. \label{DC95}\\
	&&\left\{\begin{array}{l}
	\displaystyle{w_{6y}=cw_{6x}+c_1},\vspace{2mm}\\
	\displaystyle{w_{6t}=-9(63w_{6x}^4+126w_{6x}^2w_{6x3}+63w_{6x}w_{6x2}^2+21w_{6x}w_{6x5}
		+21w_{6x2}w_{6x4}+21w_{6x3}^2+w_{6x7})c }\vspace{2mm}\\ 
	\displaystyle{\qquad +54(3w_{6x}^2+w_{6x3})c^3+27(8w_{6x}^3+8w_{6x}w_{6x3}+8c_1w_{6x}+7w_{6x2}^2+w_{6x5})c^2 +135(3w_{6x}^2}\vspace{2mm}\\ 
	\displaystyle{\qquad +w_{6x3})c_1c +\frac92w_{6x}(2c^4+45c_1^2)-27(15w_{6x}^3+15w_{6x}w_{6x3}+w_{6x5})c_1}.
	\end{array}\right. \label{DC96}
	\end{eqnarray}
	
	For the ninth-order pBKP equation \eqref{K9}, we have an analogous result for the linear superpositions of decomposition solutions.\\
	\textbf{Proposition 6.}\label{pro6}
		Let $w_1,\ w_2,\ w_3,\ w_4,\ w_5$ and $w_6$ be solutions of the ninth-order pBKP equation \eqref{K9} with the decompositions
		\begin{eqnarray}\begin{split}
		&w_{1y}=(\Phi_1+c) w_{1x},\ w_{1t}=(81\Phi_1^4+243c\Phi_1^3+243c^2\Phi_1^2
		+90c^3 \Phi_1 +9c^4) w_{1x}-c,\\
		&w_{2y}=(\Phi_2-c) w_{2x},\ w_{2t}=(81\Phi_2^4-243c\Phi_2^3+243c^2\Phi_2^2
		-90c^3 \Phi_2 +9c^4) w_{2x}-c_1,\\
		&w_{3y}=\Phi_3 w_{3x},\ w_{3t}=81\Phi_3^4 w_{3x}-c,\\
		&w_{4y}=\Phi_4w_{4x},\ w_{4t}=81\Phi_4^4w_{4x}-c_1,
		\\
		&w_{5y}=\Phi_5w_{5x}+c_1,\ w_{5t}=\bigg(81\Phi_5^4+243c_1\Phi_5^2+\frac{405}2c_1^2\bigg)w_{5x}-c,
		\\
		&w_{6y}=\Phi_6w_{6x}+c_1,\ w_{6t}=\bigg(81\Phi_6^4+243c_1\Phi_6^2+\frac{405}2c_1^2\bigg)w_{6x}-c_1,
		\\
		\end{split}\end{eqnarray}
		then $w_7=w_1+w_2,\ w_8=w_3+\frac12w_4$ and $w_9=\frac12(w_5+w_6)$ are still solutions of the ninth-order pBKP equation \eqref{K9}.

Summarizing the results thus far, we have the situation that nontrivial	for every equation in the pBKP hierarchy, there exist three possible special types of linear superposition solutions and though the solutions $w_i,\ i=1,\ \ldots,\ 6$ shown in the Propositions 1, 3 and 5 are the decomposition solutions, their special linear superposition solutions $w_7,\ w_8$ and $w_9$ are not the decomposition solutions. 
Propositions 2, 4 and 6 establish a remarkable property of the linear superpositions in nonlinear pBKP hierarchy.

\section{Multi-component decompositions of the pBKP hierarchy}

Multi-component decomposition of the pBKP hierarchy is a natural continuation of the previous work. 
The nonlinearization approach of Lax pairs or constrained flow technique by separating spatial and temporal variables makes it possible to decompose high-dimensional equations into compatible low-dimensional systems through which quite a few integrable systems are effectively obtained \cite{cbib7,cbib8,cbib9,ybib15,ybib16,xbib25}.


\subsection{Multi-component decompositions of the fifth-order pBKP equation \eqref{K5}}

The generalization of the single component decomposition described in the previous section to the multi-component case does not require any specific choice of multi-component integrable system.
As we show below, it can be defined in a general form in terms of Lax pair \eqref{Lx}-\eqref{Lt}. Imposing nonlinearization on the Lax pair \cite{xbib23}, some special solutions of the pBKP equation \eqref{K5} can be obtained simply from the compatible solutions of the coupled KdV and higher order KdV equations which is closely related to two nontrivial members in the KdV hierarchy
\begin{eqnarray}
\begin{split}\label{vyt}
v_y&=v_{x3}+av_x^2+b,\\
v_t&=9v_{x5}+30av_xv_{x3}+15av_{xx}^2+10a^2v_x^3+15cv_x+d.\\
\end{split}
\end{eqnarray}
The flows determined by \eqref{vyt} commute. Let $w$ be a solution decomposed in the form 
\begin{eqnarray}
&&w_y=F(w,w_x,w_{xx},...,w_{xm},v,v_x,v_{xx},...,v_{xn}),\\
&&w_t=G(w,w_x,w_{xx},...,w_{xr},v,v_x,v_{xx},...,v_{xs}),
\end{eqnarray}
where the compatibility condition $w_{yt}=w_{ty}$
also leads to the pBKP equation \eqref{K5}. A direct calculation shows $F$ and $G$ satisfy
\begin{eqnarray}
\begin{split}
w_y&=w_{x3}+\frac{2a^2}{3}v_x^2-2aw_xv_x+3w_x^2+c,\\
w_t&=9w_{x5}+10a^2(v_{xx}^2+2v_{x}v_{x3}+3v_x^2w_x)-30a(v_{x3}w_x+v_{xx}w_{xx}+v_xw_{x3}+3v_xw_x^2)\\
&+45(w_{xx}^2+2w_x^3+2w_xw_{x3})+15cw_x+e.
\end{split}
\end{eqnarray}
The following statement summarizes multi-composition decomposition of the fifth-order pBKP equation  \eqref{K5}.
\\
\textbf{Proposition A.}
 let $v$ and $w$ be compatible solutions of the nonlinear integrable couplings
\begin{eqnarray}\label{multide1}
\begin{split}
	v_y&=\Phi_1 v_x+b,\\
	w_y&=\Phi_2w_x+2av_x(\frac{a}{3}v_x-w_x)+c
\end{split}
\end{eqnarray}
and
\begin{eqnarray}\label{multide2}
\begin{split}
v_t&=9\Phi_1^2v_x+15cv_x+d,\\
w_t&=9\Phi_2^2w_x+10a^2(3v_x^2w_x+2v_{x3}v_x+v_{xx}^2)-30a(3v_xw_x^2+v_{x3}w_x+v_{xx}w_{xx}+v_xw_{x3})+15cw_x+e,
\end{split}
\end{eqnarray}
with $a,b,c,d$ and $e$ are arbitrary constants,
here and further on, $\Phi_1=\partial_x^2+\frac{4}{3}av_x-\frac{2}{3}a\partial_x^{-1}v_{xx},$
$\Phi_2=\partial_x^2+4w_x-2\partial_x^{-1}w_{xx},$
then, the $w$ also solves pBKP equation \eqref{K5}.

The systems determined by \eqref{multide1} and \eqref{multide2} are compatible as is easily checked. We are interested in finding the constraints such that for given decomposition solutions, their linear combinations will still be solutions to the given equation. However, linear superposition assumption appears to fail at this situation.
Introducing the new variables $v_{x}=q,w_x=p,q_{y}=q_{t},p_y=p_t$, we obtain from the Proposition A a nonlinear integrable coupled KdV equations 
\begin{eqnarray}
\begin{split}
q_t&=(q_{xx}+aq^2)_x,\\
p_t&=(p_{xx}+3p^2+\frac{2a^2}{3}q^2-2apq)_x,
\end{split}
\end{eqnarray}
which commute.
The second multi-component decomposition of the fifth-order pBKP equation \eqref{K5} is\\
\textbf{Proposition B.}
If $w$ satisfies the nonlinear integrable $(n+1)$-component KdV-type system
\begin{eqnarray}
\begin{split}
v_{iy}&=v_{ix3}+3w_xv_{ix}+b_i,\\
w_y&=\Phi_3w_{x}+\sum_{i=1}^na_iv_{ix}^2+a,\\
v_{it}&=9v_{ix5}+45(w_xv_{ixx})_x+15v_{ix}(3w_{x3}+\frac{9}{2}w_x^2+\sum_{i=1}^na_iv_{ix}^2+a)+c_i,\\
w_t&=9\Phi_3^2w_x+15aw_x+d+15\sum_{i=1}^na_i(v_{ixx}^2+2v_{ix}v_{ix3}+3w_{x}v_{ix}^2),
\end{split}
\end{eqnarray}
where $\Phi_3=\partial_x^2+2w_x-\partial^{-1}_xw_{xx}$,
then, $w$ also satisfies the pBKP equation \eqref{K5}.

Rewriting $v_{ix}=q_i,w_x=p,q_{iy}=q_{it},p_y=p_t,$ new $(n+1)$-component nonlinear integrable coupled KdV systems with $n$ sources arise
 \begin{eqnarray}
 \begin{split}\label{neqs}
 q_{it}&=(q_{ixx}+3pq_{i})_x, i=1,2,...,n,\\
 p_t&=(p_{xx}+\frac{3}{2}p^2+\sum_{i=1}^na_iq_{i}^2)_x,\\
 \end{split}
 \end{eqnarray}
 where $a_i$ are arbitrary constants.
 Multi-component coupled systems are ubiquitous in physics ranging from cold atomic systems to nonlinear optics. Such systems are typically nonlinear with considerable interactions and often have an intricate interplay between the various species \cite{cbib10,sbib39,sbib40,sbib37}.
 \\
 \textbf{System I.} The representative case $n=1$, which is related to certain types of long waves, some internal acoustic and planetary waves arising in geophysical fluid mechanics 
 	 \begin{eqnarray}
 	\begin{split}\label{sysa}
 	q_{1t}&=(q_{1xx}+3pq_{1})_x,\\
 	p_t&=(p_{xx}+\frac{3}{2}p^2+a_1q_{1}^2)_x\\
 	\end{split}
 	\end{eqnarray}
 	trivially splits into two independent KdV equations \cite{sbib38}
 	 \begin{eqnarray}\label{indekdv}
 	\begin{split}
 	w_{1t}&=(w_{1xx}+3Aw_{1}^2)_x,\\
 	w_{2t}&=(w_{2xx}+3Bw_{2}^2)_x\\
 	\end{split}
 	\end{eqnarray}
 	under a linear transformation. 
 Such a system is said to be decoupled since the variables do not interact with each other and each variable can be solved for independently.

It is to be pointed out that the two-component version \eqref{sysa} has been studied intensively in literature \cite{mbib13,sbib36,hbib3,ybib18}. Another path for deriving the System I \eqref{sysa} from a two-layer fluid system was followed by one of the authors (Lou) \cite{sbib11}. As a consequence, one function has no effect on the other and these equations are less interesting for applications. Thus, we list some genuinely non-decoupled systems that cannot be decoupled by any change of variables.\\
 \textbf{System II.} Case $n=2$, three-component KdV system with two sources is
 \begin{eqnarray}
 \begin{split}
 q_{1t}&=(q_{1xx}+3pq_{1})_x,\\
 q_{2t}&=(q_{2xx}+3pq_{2})_x,\\
 p_t&=(p_{xx}+\frac{3}{2}p^2+a_1q_{1}^2+a_2q_{2}^2)_x.
 \end{split}
 \end{eqnarray}
 	\textbf{System III.} The relevant four-component KdV system with three sources for the $n=3$ is 
 \begin{eqnarray}
 	\begin{split}
 	q_{1t}&=(q_{1xx}+3pq_{1})_x,\\
 	q_{2t}&=(q_{2xx}+3pq_{2})_x,\\
 	q_{3t}&=(q_{3xx}+3pq_{3})_x,\\
 	p_t&=(p_{xx}+\frac{3}{2}p^2+a_1q_{1}^2+a_2q_{2}^2+a_3q_{3}^2)_x.
 	\end{split}
 \end{eqnarray}

System \eqref{sysa} can be decoupled, while all the other systems are non-decoupled and seem to be new. We call these systems integrable under the meaning that they possess infinitely many higher order symmetries \cite{pbib7,sbib20,wbib21,zbib17}.

In addition to producing integrable coupled systems and their multi-component generalizations, we may use the integrable couplings to obtain new solutions of pBKP equation \eqref{K5}. Here, we extend one such example 
 \eqref{indekdv} to a more general nontrivial form.\\
\textbf{Proposition C.}
Suppose $w_1$ and $w_2$ are solutions of two independent coupled KdV systems
\begin{eqnarray}
\begin{split}\label{pc}
w_{1y}&=\Phi_4w_{1x}-\beta_1 w_{1x}+\mu,\\
w_{2y}&=\Phi_5w_{2x}+\beta_1w_{2x}+\mu_1,\\
w_{1t}&=9\Phi_4^2w_{1x}-15\beta_1\Phi_4w_{1x}+5(\beta_1^2+3c_1\mu+3c_2\mu_1)w_{1x}+\delta,\\
w_{2t}&=9\Phi_5^2w_{2x}+15\beta_1\Phi_5w_{2x}+5(\beta_1^2+3c_1\mu+3c_2\mu_1)w_{2x}+\delta_1
\end{split}
\end{eqnarray}
with
$\Phi_4=\partial_x^2+4c_1w_{1x}-2c_1\partial_{x}^{-1}w_{1xx}$,
$\Phi_5=\partial_x^2+4c_2w_{2x}-2c_2\partial_{x}^{-1}w_{2xx}$,
then, a linear combination $w=c_1w_1+c_2w_2$ is satisfying the pBKP equation \eqref{K5}.

One can easily verify that pBKP equation \eqref{K5} holds whenever $w_1$ and $w_2$ are solutions of \eqref{pc}.
Proposition C makes it possible for us to construct different types of linear superposition solutions. In particular, let us consider a superposition of $n$ solitons and $m$ solitons with \cite{kbib1,sbib31,rbib19}
\begin{eqnarray*}
w_1=\frac{2}{c_1}\ln(f_n)_x, ~~w_2=\frac{2}{c_2}\ln(g_m)_x,
\end{eqnarray*}
where
\begin{eqnarray*}
&&f_n=\sum_{\alpha=0,1}\exp(\sum_{i=1}^n\alpha_i\xi_i+\sum_{1\leq i<j\leq n}\alpha_i\alpha_j\theta_{ij}),~g_m=\sum_{\alpha=0,1}\exp(\sum_{i=1}^m\alpha_i\eta_i+\sum_{1\leq i<j\leq m}\alpha_i\alpha_j\tau_{ij}),\\
&&\xi_i=k_ix+(k_i^3-\beta_1k_i)y+l_iz+(9k_i^5-15k_i^3\beta_1+5k_i\beta_1^2)t+\xi_{i0}, \exp(\theta_{ij})=\displaystyle\frac{(k_i-k_j)^2}{(k_i+k_j)^2},\\
&&\eta_i=m_ix+(m_i^3+\beta_1m_i)y+n_iz+(9m_i^5+15m_i^3\beta_1+5m_i\beta_1^2)t+\eta_{i0}, \exp(\tau_{ij})=\displaystyle\frac{(m_i-m_j)^2}{(m_i+m_j)^2},\\
&&\mu=0,\mu_1=0,\delta=0,\delta_1=0.
\end{eqnarray*}
The summation of $\alpha$ should be done for all permutations of $\alpha_i=0, i=1,2,....$
All the parameters
$c_i,c_2,\beta_1,k_i,l_i,m_i,n_i,\xi_{i0}$ and $\eta_{i0}$ are arbitrary constants.

We conclude this subsection with \textbf{Remarks}. 1.
The $(n+1)$-component nonlinear integrable coupled KdV systems
\eqref{neqs} with $n$ sources are non-decouplable. A decoupled system only appears in the case of $n=1$. It is worth emphasizing that a natural and meaningful consequence of decoupled system I is any KdV equation cannot solve the pBKP equation \eqref{K5}, whereas the sum of two arbitrary KdV equations does.
More precisely, neither $w = c_1w_1$ nor $w = c_2w_2$ in Proposition C is a solution of pBKP equation \eqref{K5} to any $c_1$ and $c_2$, but $w = c_1w_1 + c_2w_2$ is. The possible connection between integrable systems is established utilizing linear superposition.
2. Both the first two decompositions in Proposition 1 and the known linear superpositions in Proposition 2 are included in Proposition C with specific $c_1$ and $c_2$.

\subsection{Multi-component decompositions of the seventh-order pBKP equation \eqref{K7}}

In close analogy with the calculation of the multi-component decompositions of the fifth-order pBKP equation \eqref{K5}, the following similar statements hold for the seventh-order pBKP equation \eqref{K7}. \\
\textbf{Proposition D.}
Suppose $w$ satisfies the nonlinear integrable couplings
\begin{eqnarray}\label{sede1}
\begin{split}
v_y&=\Phi_1 v_x+b,\\
w_y&=\Phi_2w_x+2av_x(\frac{a}{3}v_x-w_x)+c
\end{split}
\end{eqnarray}
and
\begin{eqnarray}\label{sede2}
\begin{split}
v_t&=27\Phi_1^3v_x+63c\Phi_1v_x+dv_x+f,\\
w_t&=27\Phi_2^3w_x-126a\Phi_2^2w_x+63c\Phi_2w_x+\frac{70}{3}a^4v_x^4-140a^3v_x^3w_x\\&+42a^2(15v_x^2w_x^2+10w_{xx}v_{xx}v_x+5w_{x3}v_x^2+5w_xv_{xx}^2+10v_{x3}w_xv_x+cv_x^2+4v_{xx}v_{x4}+2v_xv_{x5}+3v_{x3}^2)\\&-126a(10v_{xx}w_xw_{xx}+5v_{x3}w_x^2+cv_xw_x+2w_{xx}v_{x4}+3w_{x3}v_{x3}+w_xv_{x5}+2w_{x4}v_{xx})+dw_x+e, 
\end{split}
\end{eqnarray}
then, $w$ also satisfies the seventh-order pBKP equation \eqref{K7}.
\\
\textbf{Proposition E.}
If $w$ is a solution of the nonlinear integrable $(n+1)$-component KdV system
\begin{eqnarray}
\begin{split}
v_{iy}&=v_{ix3}+3w_xv_{ix}+b_i,\\
w_y&=\Phi_3w_{x}+\sum_{i=1}^na_iv_{ix}^2+a,\\
v_{it}&=27v_{ix7}+189(v_{ix5}w_x+v_{ix}w_{x5})+378(v_{ix4}w_{xx}+v_{ixx}w_{x4})+945(v_{ix}w_xw_{x3}\\&+v_{ixx}w_xw_{xx})+\frac{945}{2}(w_x^2v_{ix3}+v_{ix}w_{xx}^2+w_x^3v_{ix})+567v_{ix3}w_{x3}+63av_{ix3}\\&+189av_{ix}w_x+bv_{ix}+c_i+315a_iv_{ix}(v_{ix}v_{ix3}+v_{ixx}^2+v_{ix}^2w_x)\\&+63\sum_{j\neq i}^n a_j(v_{ix3}v_{jx}^2+2v_{ixx}v_{jx}v_{jxx}+3v_{ix}v_{jxx}^2+4v_{ix}v_{jx}v_{jx3}+5v_{ix}v_{jx}^2w_x),\\
w_t&=27\Phi_3^3w_x+63a\Phi_3w_x+bw_x+d+\frac{105}{2}\sum_{i=1}^na_i^2v_{ix}^4\\
&+63\sum_{i=1}^n(2v_{ix}v_{ix5}+4v_{ixx}v_{ix4}+3v_{ix3}^2+av_{ix}^2+5v_{ix}^2w_{x3}+10v_{ix}v_{ixx}w_{xx}\\
&+5(2v_{ix}v_{ix3}+v_{ixx}^2)w_x+\frac{15}{2}v_{ix}^2w_x^2)+105\sum_{i<j}^na_ia_jv_{ix}^2v_{jx}^2,
\end{split}
\end{eqnarray}
then, $w$ is also a solution of the seventh-order pBKP equation \eqref{K7}.
\\
\textbf{Proposition F.}
If $w_1$ and $w_2$ are compatible solutions of two independent KdV systems
\begin{eqnarray}
\begin{split}
w_{1y}&=\Phi_4w_{1x}-\beta_1 w_{1x}+\mu,\\
w_{2y}&=\Phi_5w_{2x}+\beta_1w_{2x}+\mu_1,\\
w_{1t}&=27\Phi_4^3w_{1x}-63\beta_1\Phi_4^2w_{1x}+(42\beta_1^2+63c_1\mu+63c_2\mu_1)\Phi_4w_{1x}-(7\beta_1^3+126c_1\mu\beta_1)w_{1x}+\delta,
\\
w_{2t}&=27\Phi_5^3w_{2x}+63\beta_1\Phi_5^2w_{2x}+(42\beta_1^2+63c_1\mu+63c_2\mu_1)\Phi_5w_{2x}+(7\beta_1^3+126c_2\mu_1\beta_1)w_{2x}+\delta_1
\end{split}
\end{eqnarray}
then, $w=c_1w_1+c_2w_2$ solves the seventh-order pBKP equation \eqref{K7} exactly.

Remarkably, the first two decompositions in Proposition 3 and the known linear superpositions in Proposition 4 are special cases of Proposition F with specific $c_1$ and $c_2$.

\subsection{Multi-component decompositions of the ninth-order pBKP equation \eqref{K9}}

The results in this subsection are similar to the previous ones.\\
	\textbf{Proposition G.}
	If $w$ is a solution of the nonlinear integrable couplings
	\begin{eqnarray}\label{nide1}
	\begin{split}
	v_y&=\Phi_1 v_x+b,\\
	w_y&=\Phi_2w_x+2av_x(\frac{a}{3}v_x-w_x)+c
	\end{split}
	\end{eqnarray}
	and
	\begin{eqnarray}\label{nide2}
	\begin{split}
	v_t&=81\Phi_1^4v_x+243c\Phi_1^2v_x+3\Phi_1v_x+dv_x+e,\\
	w_t&=81\Phi_2^4w_x-486av_x\Phi_2^3w_x+81(3c+14a^2v_x^2)\Phi_2^2w_x-3(420a^3v_x^3+270acv_x-1)\Phi_2w_x\\
	&-486aw_x\Phi_1^3v_x+420a^4(3v_x^4w_x+2v_x^3v_{x3}+3v_x^2v_{xx}^2)a^4-3780a^3w_{xx}v_x^2v_{xx}+2a^2(162v_xv_{x7}\\
	&+486v_{xx}v_{x6}+54(42v_xw_x+19v_{x3})v_{x5}+621v_{x4}^2+2268(v_xw_{xx}+2w_xv_{xx})v_{x4}\\&+2268v_xv_{xx}w_{x4}+3402w_xv_{x3}^2+54(63v_xw_{x3}+77w_{xx}v_{xx}+(105w_x^2+5c)v_x)v_{x3}\\&+2079v_{xx}^2w_{x3}+135(21w_x^2+c)v_{xx}^2+11340v_xw_xv_{xx}w_{xx}+405cw_xv_x^2+v_x^2)\\&-6a(243w_{xx}v_{x6}+243v_{xx}w_{x6}+27(21w_x^2+19w_{x3})v_{x5}+513v_{x3}w_{x5}+27(84w_xw_{xx}\\&+23w_{x4})v_{4x}+2268v_{xx}w_xw_{x4}+27(70w_x^3+126w_xw_{x3}+5cw_x+77w_{xx}^2)v_{x3}\\&+4158v_{xx}w_{xx}w_{x3}+135(42w_x^2+c)w_{xx}v_{xx}+v_xw_x)+dw_x+f,
	\end{split}
	\end{eqnarray}
	then, $w$ is also a solution of the ninth-order pBKP equation  \eqref{K9}.
	\\
\textbf{Proposition H.}
Let $v$ and $w$ be compatible solutions of the nonlinear coupled KdV system
\begin{eqnarray}
\begin{split}
v_{y}&=v_{x3}+3w_xv_{x}+b_1,\\
w_y&=\Phi_3w_{x}+a_1v_{x}^2+a,\\
v_t&=81v_{x9}+243(3v_{x7}w_x+3v_xw_{x7}+9v_{x6}w_{xx}+9v_{xx}w_{x6}+19v_{x5}w_{x3}+19v_{x3}w_{x5}+23v_{x4}w_{x4})\\
&+1701(3v_xw_xw_{x5}+6v_xw_{xx}w_{x4}+6w_xv_{xx}v_{x4}+6w_xv_{x4}w_{xx}+9w_xv_{x3}w_{x3}+11v_{xx}w_{xx}w_{x3})\\
&+\frac{1701}{2}(3w_x^2v_{x5}+5v_{x3}w_x^3+9v_xw_{x3}^2+11w_{xx}^2v_{x3}+\frac{15}{4}v_xw_x^4)+e(v_{x3}+3v_xw_x)\\&+\frac{25515}{2}(v_{xx}w_x^2w_{xx}+v_xw_xw_{xx}^2+v_xw_x^2w_{x3})+bv_x+\frac{567}{2}a_1^2v_x^5+567a_1(3v_x^2v_{x5}+5v_x^3w_{x3}\\&+9v_xv_{x3}^2+11v_{x3}v_{xx}^2+12v_xv_{xx}v_{x4}+15v_x^2v_{x3}w_x+15v_x^2v_{xx}w_{xx}+15v_xv_{xx}^2w_x+\frac{15}{2}v_x^3w_x^2)\\&+81a(3v_{x5}+5a_1v_x^3+15v_{x3}w_x+15v_{xx}w_{xx}+15v_xw_{x3}+\frac{45}{2}v_xw_x^2),\\
w_t&=81\Phi_3^4w_x+243a\Phi_3^2w_x+e\Phi_3 w_x+bw_x+d+\frac{945}{2}a_1^2v_x^2(3v_x^2w_x+4v_xv_{x3}+6v_{xx}^2)\\&+a_1[81(6v_xv_{x7}+18v_{xx}v_{x6}+38v_{x3}v_{x5}+23v_{x4}^2)+\frac{8505}{2}(v_x^2w_x^3+v_x^2w_{xx}^2+v_{xx}^2w_x^2)\\
&+8505(v_x^2w_xw_{x3}+v_xv_{x3}w_x^2)+567(3v_x^2w_{x5}+9v_{x3}^2w_x+11v_{xx}^2w_{x3})\\&+1134(3v_xv_{x5}w_x+6v_xv_{xx}w_{x4}+6v_xv_{x4}w_{xx}+6v_{xx}v_{x4}w_x+9v_xv_{x3}w_{x3}+11v_{xx}v_{x3}w_{xx})\\&+17010v_xv_{xx}w_xw_{xx}+405a(v_{xx}^2+2v_xv_{x3}+3v_x^2w_x)+ev_x^2],
\end{split}
\end{eqnarray}
then, $w$ solves the seventh-order pBKP equation \eqref{K9} as well.

\textbf{Proposition I.}
	If $w_1$ and $w_2$ are solutions of two independent KdV equations
	\begin{eqnarray}
	\begin{split}
	w_{1y}&=\Phi_4w_{1x}-\beta_1 w_{1x}+\mu,\\
	w_{2y}&=\Phi_5w_{2x}+\beta_1w_{2x}+\mu_1,\\
	w_{1t}&=81\Phi_4^4w_{1x}-243\beta_1\Phi_4^3w_{1x}+243(c_1\mu+c_2\mu_1+\beta_1^2)\Phi_4^2w_{1x}-90\beta_1(\beta_1^2+9c_1\mu)\Phi_4w_{1x}\\&-810c_2\mu_1\beta_1^2w_{1x}+\delta,\\
	w_{2t}&=81\Phi_5^4w_{2x}+243\beta_1\Phi_5^3w_{2x}+243(c_1\mu+c_2\mu_1+\beta_1^2)\Phi_5^2w_{2x}+90\beta_1(\beta_1^2+9c_2\mu_1)\Phi_5w_{2x}\\&-810c_1\mu \beta_1^2w_{2x}+\delta_1,
	\end{split}
	\end{eqnarray}
	then, $w=c_1w_1+c_2w_2$ is a solution of the  ninth-order pBKP equation \eqref{K9}.

Note the first two decompositions in Proposition 5 and the known linear superpositions in Proposition 6 are special cases of Proposition I with specific $c_1$ and $c_2$.
All propositions in this paper hold through direct calculation.
It is natural to generalize the decompositions used in this paper to the whole pBKP hierarchy. Extending the argument, we conjecture that for each equation in the pBKP hierarchy, there exist multi-component decompositions and a special general linear superposition.


\section{Conclusions and discussions}

This paper motivates the study of multi-component decompositions and linear superpositions of nonlinear pBKP hierarchy. We mainly deal with novel multi-component decompositions of three pBKP equations by enlarging spectral problem and as such these decompositions allow us to derive certain new nonlinear integrable coupled KdV-type systems with $n$ sources which possess infinitely many higher order symmetries and construct a general linear superposition solution that is related to two arbitrary KdV couplings. 
It should be noted that the linear superpositions in Propositions 2, 4 and 6 are special cases of Propositions C, F and I with specific parameters.

Under the framework of this paper, it is hoped that a wealth of integrable multi-component decompositions for other high-dimensional nonlinear systems can be revealed. 
Another interesting direction is to generalize decomposition to consider more general multi-component decompositions.
At present, the expressions and methods of linear superposition in nonlinear theory are still in a very preliminary state, there is no general statement about linear superposition of solutions.
We expect to see more examples of nonlinear equations that their solutions satisfy the principle of linear superposition, and, finally, a systematic theory. In addition, although the integrable couplings are obtained by multi-composition decompositions, the decomposition process does not provide any solution to the new integrable couplings. It is still a question how to solve the integrable couplings.

\section*{Acknowledgement}
The work was sponsored by the National Natural Science Foundations of China (Nos. 11975131, 11435005), K. C. Wong Magna Fund in Ningbo University, the Natural Science Foundation of Zhejiang Province No. LQ20A010009 and the General Scientific Research of Zhejiang Province No. Y201941009.

\bibliographystyle{elsarticle-num}
\bibliography{ref}

\end{document}